\documentclass[journal,twoside,web]{ieeecolor}
\def\logoname{LOGO-blank}
\definecolor{subsectioncolor}{rgb}{0,0.541,0.855}
\usepackage{cite}
\usepackage{amsmath,amssymb,amsfonts}
\usepackage{graphicx}
\usepackage{textcomp}
\usepackage[utf8]{inputenc} 
\usepackage[T1]{fontenc}    
\usepackage{url}            
\usepackage{booktabs}       
\usepackage{nicefrac}       
\usepackage{microtype}      
\usepackage{subcaption}     
\usepackage[font={small}]{caption}
\usepackage{siunitx}
\usepackage{xspace}
\newcommand{\ie}{i.e.\@\xspace}
\newcommand{\invivo}{\textit{in vivo}\xspace}

\begin{document}  
\title{SweiNet: Deep Learning Based Uncertainty Quantification for Ultrasound Shear Wave Elasticity Imaging}
\author{Felix Q. Jin$^1$, 
Lindsey C. Carlson$^{2,3}$,
Helen Feltovich$^3$,
Timothy J. Hall$^2$,
Mark L. Palmeri$^1$
\\
\vspace{1ex}
$^1$ Department of Biomedical Engineering, Duke University, Durham, NC\\
$^2$ Department of Medical Physics, University of Wisconsin, Madison, WI\\
$^3$ Maternal Fetal Medicine, Intermountain Healthcare, Provo, UT
\thanks{
Corresponding author: Felix Jin (e-mail: \url{felix.jin@duke.edu}).
Trained model and code available at \url{https://github.com/fqjin/swei-net}.
}
}

\maketitle

\begin{abstract}  
In ultrasound shear wave elasticity (SWE) imaging, a number of algorithms exist for estimating the shear wave speed (SWS) from spatiotemporal displacement data.
However, no method provides a well-calibrated and practical uncertainty metric, hindering SWE's clinical adoption and utility in downstream decision-making.
Here, we designed a deep learning SWS estimator that simultaneously outputs a quantitative and well-calibrated uncertainty value for each estimate.
Our deep neural network (DNN) takes as input a single 2D spatiotemporal plane of tracked displacement data and outputs the two parameters $m$ and $\sigma$ of a log-normal probability distribution.
For training and testing, we used \invivo 2D-SWE data of the cervix collected from 30 pregnant subjects, totaling 551 acquisitions and \textgreater2 million space-time plots.
Points were grouped by uncertainty into bins to assess uncertainty calibration: the predicted uncertainty closely matched the root-mean-square estimation error, with an average absolute percent deviation of 3.84\%.
We created a leave-one-out ensemble model that estimated uncertainty with better calibration (1.45\%) than any individual ensemble member on a held-out patient's data.
Lastly, we applied the DNN to an external dataset to evaluate its generalizability.
We have made the trained model, SweiNet, openly available to provide the research community with a fast SWS estimator that also outputs a well-calibrated estimate of the predictive uncertainty.
\end{abstract}


\section{Introduction}
\IEEEPARstart{U}{ltrasound} shear wave elasticity (SWE) imaging, also called SWEI, is a non-invasive technique that evaluates tissue elastic properties by measuring the speed of shear waves~\cite{sarvazyan1998shear,sarvazyan2011overview}.
Micron-scale shear waves are generated via an acoustic radiation force impulse (``push'')~\cite{nightingale2003shear};
these shear waves are polarized axially and propagate outwards laterally.
Ultrafast ultrasonic imaging with correlation-based displacement estimators~\cite{kasai1985real,loupas1995axial,pinton2006rapid} tracks particle motion, from which the shear wave speed (SWS) is estimated.
In a linear, elastic, homogeneous and isotropic material, the SWS is related to the shear modulus $G$ by the equation $G = \rho \, c^2$, where $\rho$ is the tissue density and $c$ is the SWS.
Furthermore, if the material is incompressible, then the Young's modulus is equal to $3G$.
Under these typical assumptions, tissue stiffness can be directly quantified by measuring the SWS.

\begin{figure}[t]
	\centering
	\includegraphics[width=\columnwidth]{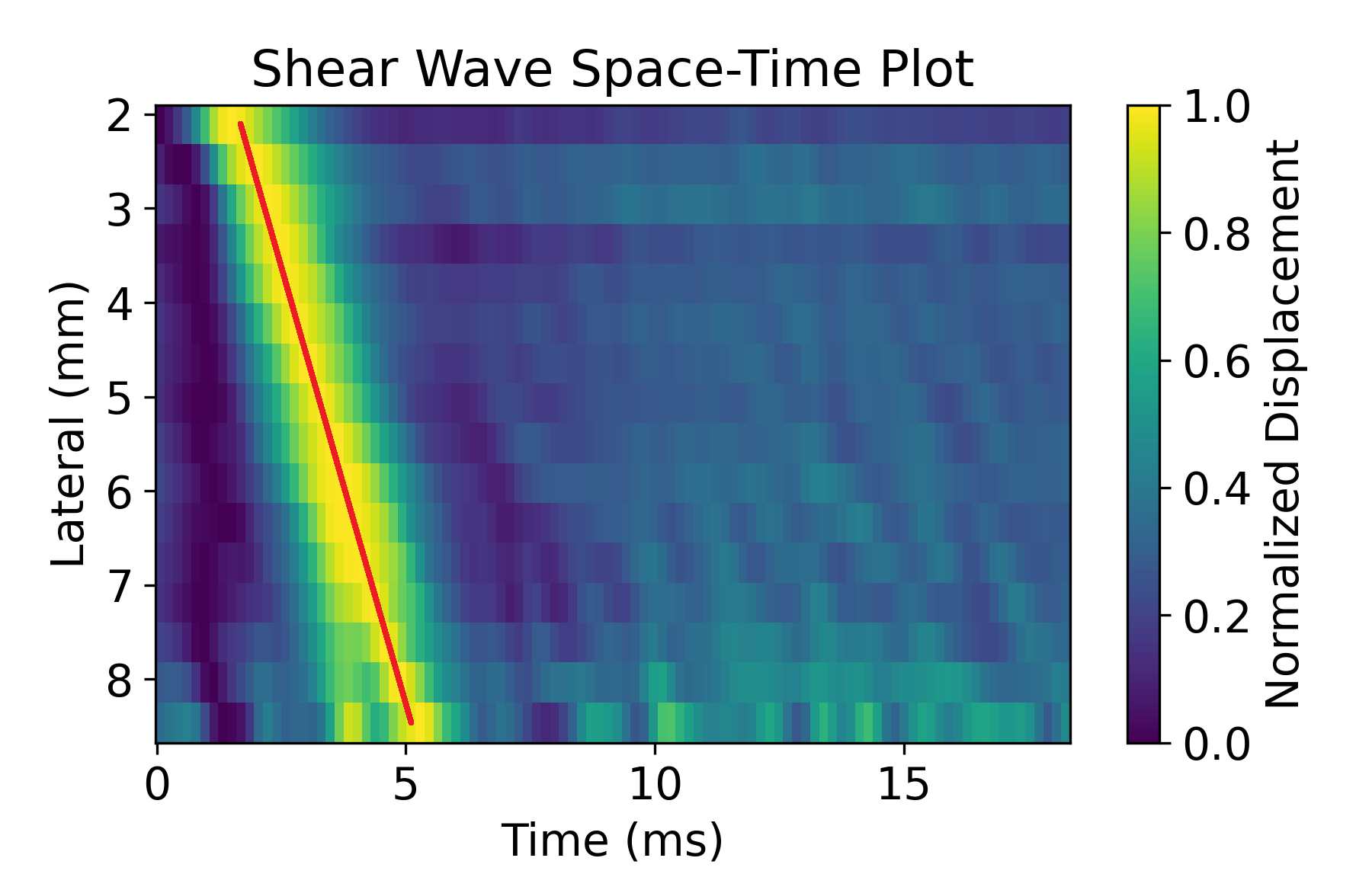}
	\caption{
	Space-time plot of shear wave data from the \invivo cervix dataset.
	The slope of the apparent wavefront (red line) corresponds to the group shear wave speed, estimated to be 2.0 m/s.
	}
	\label{fig:spacetime}
\end{figure}
\subsection{SWS Estimation}
A useful way to visualize shear wave propagation is with the space-time plot, where tracked displacement is plotted over lateral positions and times (Fig.~\ref{fig:spacetime}).
A shear wave with constant speed will appear as a linear wavefront, with the slope corresponding to the group SWS.
The central task addressed in this paper is the estimation of the SWS (\ie, the slope) from a given space-time plot.

\textit{In vivo} data pose a significant challenge for SWS estimation due to tissue heterogeneity, patient motion, acoustic clutter, reflection artifacts, jitter noise, and speckle-associated bias~\cite{wang2010improving,rouze2012parameters,mcaleavey2015shear,deng2016system}.
A number of SWS-estimation algorithms have been developed, and they can be broadly divided into two groups: time-of-flight (TOF) and transform-based methods.

TOF methods use differences in wave arrival time to estimate the SWS.
The simplest approach is to calculate the time to peak particle displacement at each track location and use a linear regression to find the slope~\cite{palmeri2008quantifying}.
More sophisticated forms of linear regression like the RANSAC algorithm can reduce sensitivity to gross outliers~\cite{wang2010improving}.
Time-to-peak is easy to calculate but throws away information outside the peak.
Cross-correlation of data between adjacent track locations leverages the entire waveform and can provide a better estimate of the travel time difference~\cite{tanter2008quantitative,song2014fast}.

The second group of methods use an integral transform to view the displacement data in a different domain where the wave speed can be readily determined.
The Radon transform computes a line integral of the space-time plot for all trajectories.
The peak of the Radon transform corresponds to the wave's trajectory, and the slope of the line is the SWS~\cite{rouze2010robust,urban2012use,jin2022radon}. 
The Fourier-transform is another transform and can be used to separate data by spatial and temporal frequency to find the local phase velocity~\cite{kijanka2018local}.

Displacement data is sometimes differentiated to generate particle velocity or acceleration data, which can help suppress motion artifacts but alters the wave's frequency content~\cite{deng2016ultrasonic,rouze2018characterization}.
SWS-estimation algorithms typically work with both particle displacement and velocity data.

Measurement uncertainty is useful in clinical decision making, yet only a few SWS estimators output an associated uncertainty metric.
None of these is well-calibrated or easily interpreted by a clinician.
Most are numbers between 0 and 1 and do not have units in meters per second.
Common metrics include $R^2$, the percentage of inliers (RANSAC), and the correlation coefficient between adjacent track locations.
Transform-based methods do not have a well-defined uncertainty metric. Specifically, the width of the peak in the Radon transform is generally not predictive of the likelihood of measurement error.
Some clinical scanners will report the standard deviation of measured SWSs over a region of interest (ROI).
While this value is interpretable and has the right units, it describes observed spatial variation and not the measurement uncertainty at any given point.

\subsection{Deep Learning \& Uncertainty Quantification}
Deep learning is a powerful and versatile machine learning tool that has outperformed explicitly-designed algorithms in a wide number of fields~\cite{lecun2015deep}.
Deep learning has previously been applied to ultrasound beamforming~\cite{luchies2018deep,hyun2019beamforming} and ultrasonic displacement tracking~\cite{chan2021deep,tehrani2020displacement}.
Deep learning has the potential to improve ultrasound SWE by performing SWS estimation better and faster, but existing research is lacking.
One peer-reviewed work was relatively limited in scope, used phantom and simulated data, and compared their results only to a phase-velocity-based algorithm~\cite{ahmed2020dswe}.

The majority of deep learning applications take model outputs blindly without regard to uncertainty or confidence, yet models often produce overconfident predictions, even in situations where they are completely wrong, such as with random inputs~\cite{hein2019relu,nguyen2015deep,ovadia2019can}.
Various techniques to quantify uncertainty have been proposed, including direct variance estimation~\cite{nix1994estimating,kendall2017uncertainties}, dropout variational inference~\cite{gal2016dropout}, ensembles~\cite{lakshminarayanan2017simple}, and stochastic gradient Markov chain Monte Carlo~\cite{li2016learning,zhang2019cyclical}.

For direct variance estimation, one trains a deep neural network (DNN) to output the parameters of a probability distribution $p(y)$ for the target variable $y$~\cite{nix1994estimating,kendall2017uncertainties}.
These authors used a normal (Gaussian)  distribution 
\begin{equation}
    p(y|\mu,\sigma) = 
    \frac{1}{\sqrt{2\pi\sigma^2}} 
    \exp\left(\frac{-(y-\mu)^2}{2\sigma^2}\right) 
    \,,
\end{equation}
where the location parameter $\mu$ represents the estimated value and the shape parameter $\sigma$ represents the predictive uncertainty, \ie how far the estimated value is likely to be from the true value.
Given data, the network is trained to minimize the negative log-likelihood, which is equivalent to maximum likelihood estimation.
The appropriate loss function is
\begin{equation}\label{eq:normal_loss}
    \mathcal{L} = 
    \frac{(y-\mu)^2}{\sigma^2} 
    + \log \sigma^2
    \,. 
\end{equation}
When $\sigma$ is held constant, the loss just simplifies to the mean squared error.
In equation~\ref{eq:normal_loss} the squared error is inversely weighted by the estimated variance so that samples with higher uncertainty contribute less to the loss.
The second term monotonically increases with $\sigma$ and acts to prevent $\sigma$ from becoming too large.
If the true value $y$ is drawn from a normal distribution, then the parameters $\mu^*$ and $\sigma^*$ that minimize the loss function are exactly the mean and standard deviation of the distribution.
The model learns a calibrated uncertainty implicitly during training without an explicit uncertainty target.
Unlike other methods, this approach does not require sampling at inference or running multiple models.

\begin{figure*}[t]
	\centering
	\includegraphics[width=\textwidth]{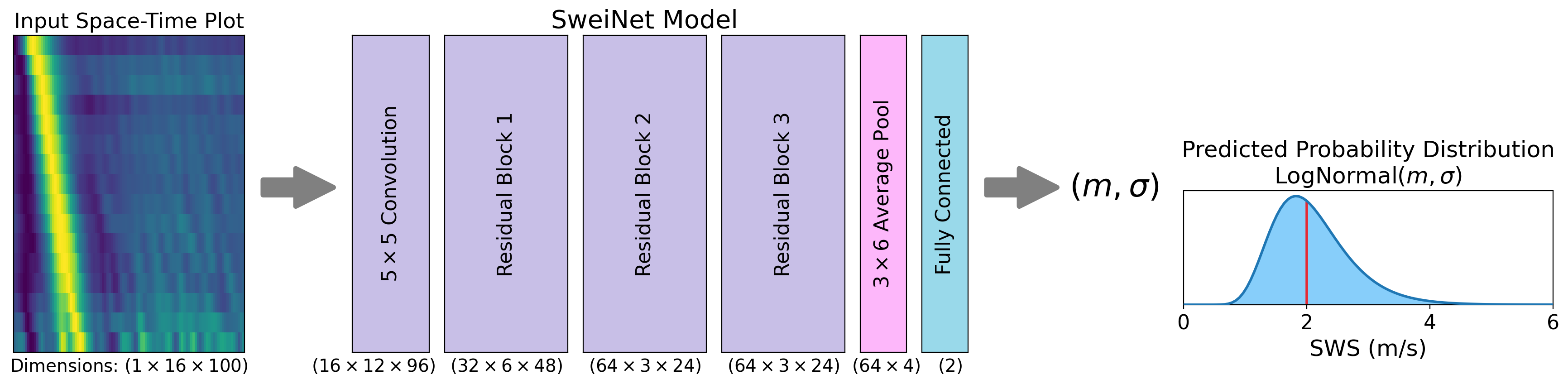}
	\caption{
	Deep neural network architecture.
	The input is a single space-time plot of shear wave data, and the outputs are two quantities $m$ and $\sigma$ that parameterize a log-normal likelihood distribution.
	The example log-normal distribution has parameters $(2,0.3)$, and the median $m$ is indicated by the red line.
	}
	\label{fig:architecture}
\end{figure*}

\subsection{Cervix Elastography in Pregnancy}
The cervix is the lower part of the uterus that acts as the gateway to delivery, and remodels dramatically throughout pregnancy from firm, long and closed as the fetus develops \textit{in utero} to short, soft, and dilated to allow delivery~\cite{timmons2010cervical}.
In obstetrics, predicting the risk of spontaneous preterm birth or the success of induction of labor are challenging problems that currently rely on many factors including a clinician's subjective assessment of cervical softening~\cite{feltovich2019labour}.
Ultrasound SWE has the potential to address this clinical need by providing objective and quantitative measurements of cervix stiffness~\cite{feltovich2019labour,carlson2019quantitative,swiatkowska2011elastography,wang2019diagnostic,lu2020predictive}.
However, existing SWS estimators are slow in practice and lack a well-calibrated uncertainty metric, hindering clinical adoption of ultrasound SWE technology.

In this paper, we use a large set of \invivo cervix SWE data to train a DNN to perform SWS estimation and uncertainty quantification.
Section~\ref{sec:methods} describes the SWE data used in this study, why we used the log-normal distribution for uncertainty quantification, our DNN architecture and training, and how we evaluated whether the predicted uncertainty was calibrated.
Section~\ref{sec:results} describes the results of leave-one-out (LOO) cross-validation and the performance of an ensemble on held-out test data.
We also applied the trained DNN on an external dataset collected in an elasticity phantom with a different transducer and scanner, demonstrating the generalizability of our trained network.
These results are followed by a discussion (Section~\ref{sec:discussion}) and conclusion.

\section{Methods}\label{sec:methods}
\subsection{SWE Data}
Ultrasound SWE data of the cervix were collected in 30 pregnant nulliparous patients at 18--22 weeks as part of an ongoing IRB-approved clinical study at Intermountain Healthcare (Provo, UT).
Using a Siemens ACUSON Sequoia scanner and an 18H6 endovaginal linear array transducer, two sonographers each acquired about nine acquisitions per patient.

In our previous study~\cite{jin2021deep}, we used point-SWE data (``VTQ''~/~pSWE), where each acquisition consisted of two push events near a single point that generate data for one left- and right-moving shear wave.
In contrast, this study used 2D SWE (``VTIQ''), where each acquisition consisted of 27 push events spaced 0.85 mm apart, mapping over the 2D ROI.
The push had a frequency of \SI{7.3}{\mega\hertz}, a duration of \SI{500}{\micro\second}, and an approximately F/3.5 focal geometry.

Pulse-inversion tracking~\cite{doherty2013harmonic} was used with a transmit frequency of \SI{7}{\mega\hertz}.
Tracking PRF was \SI{5.56}{\kilo\hertz}.
Particle displacement was computed using the Loupas algorithm~\cite{loupas1995axial,pinton2006rapid} with a \SI{1.6}{\milli\meter} axial kernel and 90\% overlap.
A quadratic fit motion filter, 60-1200 Hz second order IIR bandpass filter, and directional filter were applied to the displacement data~\cite{deng2016ultrasonic,lipman2016evaluating}.
Displacements measured at each track location were normalized independently over time to be between 0 and 1.
The cross-correlation TOF and Radon transform~\cite{jin2022radon} SWS estimators were used.
We did not prefer one method over the other, so a random value between the two methods' estimates was used as ground truth.

Data from one of the 30 patients had only 26 push events per acquisition and was set aside as a held-out test case.
We performed LOO cross-validation with the remaining 29 patients.

\subsection{Quantitative Uncertainty Estimation}
The normal distribution is not ideal for SWS estimation for a number of reasons: it has non-zero probability for negative speeds, its inverse distribution is not well-defined, and the square of a Gaussian variable is not Gaussian (it is a chi-square distribution).
The log-normal distribution is more ideal:
\begin{equation}
    p(y|m,\sigma) = \frac{1}{y\sqrt{2\pi\sigma^2}} \exp\left( \frac{-(\log{y}-\log{m})^2}{2\sigma^2} \right) \,,
\end{equation}
where $\log{m}$ is the log-domain mean and $\sigma$ is the log-domain standard deviation.
The quantity $m$ is also the median.
The region of support is positive only, and variables remain log-normal under scaling and power transformations.
Importantly, if the SWS $c$ is log-normal with parameters $m$ and $\sigma$, then the shear modulus $G = \rho \, c^2$ is also log-normal with parameters $\rho \, m^2$ and $2\sigma$.

The negative log likelihood of the log-normal distribution is
\begin{equation}
    -\log{p(y|m,\sigma)} = \frac{(\log{y}-\log{m})^2}{2\sigma^2} + \frac{1}{2} \log{2\pi\sigma^2} + \log{y} \,,
\end{equation}
which reduces to the following loss function:
\begin{equation}\label{eq:lognorm_loss}
\mathcal{L} = \frac{(\log{y}-\log{m})^2}{\sigma^2} + \log{\sigma^2} \,.
\end{equation}
This is similar to equation~\ref{eq:normal_loss} except the regular squared error is replaced by the log-domain squared error.
The estimated SWS in meters per second is $m$.
The quantity $\sinh{\sigma}$ is the unitless relative uncertainty, analogous to a coefficient of variation.
To write down an uncertainty with units of meters per second, we use the expression $m \sinh{\sigma}$.

As in~\cite{kendall2017uncertainties}, we actually train the network to output two unconstrained real values $\mu = \log{m}$ and $s = \log{\sigma^2}$.
This can be thought of as a final activation layer converting two real-valued outputs to the positive-valued $m$ and $\sigma$.

\subsection{Network Architecture and Training}
Fig.~\ref{fig:architecture} shows a schematic of the network architecture.
The convolutional DNN takes a single 2D spacetime plot as input and outputs  $m$ and $\sigma$.
The network consisted of an un-padded ${5\times5}$ input convolution layer, three pre-activation residual blocks~\cite{he2016identity}, ${3\times6}$ average pooling, and a final fully-connected layer.
All activations were LeakyReLU.
To ensure that our model could run in real time, we used a small network size with $\sim$100,000 trainable parameters.

The DNN was trained using equation~\ref{eq:lognorm_loss} and the Adam optimizer with batch size of 128, learning rate of 5e-4, weight decay of 1e-4, the 1cycle learning rate scheduler~\cite{smith2019super}, for 90 epochs.
Networks were implemented in PyTorch 1.8 and trained using an NVIDIA V100 GPU.
A trained model and the code are available at \url{https://github.com/fqjin/swei-net}.

\subsection{Evaluating Uncertainty Calibration}
A well-calibrated uncertainty value is one that closely approximates the true spread in the estimation error.
To quantify uncertainty calibration, we grouped points into 25 bins by increasing relative uncertainty and compared each bin's average relative uncertainty versus root-mean-square (RMS) relative error.
The relative scale is most natural for the log-normal distribution, and it eliminates the effect of different absolute SWSs, as points with greater SWS will tend to have greater uncertainty, all else being equal~\cite{wang2013precision}.
We can summarize the calibration curve by computing the absolute percent deviation between the predicted uncertainty and the true RMS error, averaged over the 25 bins.
This quantitatively measures how well the DNN model predicts the probability of estimation error.

\begin{figure}[t]
	\centering
	\includegraphics[width=\columnwidth]{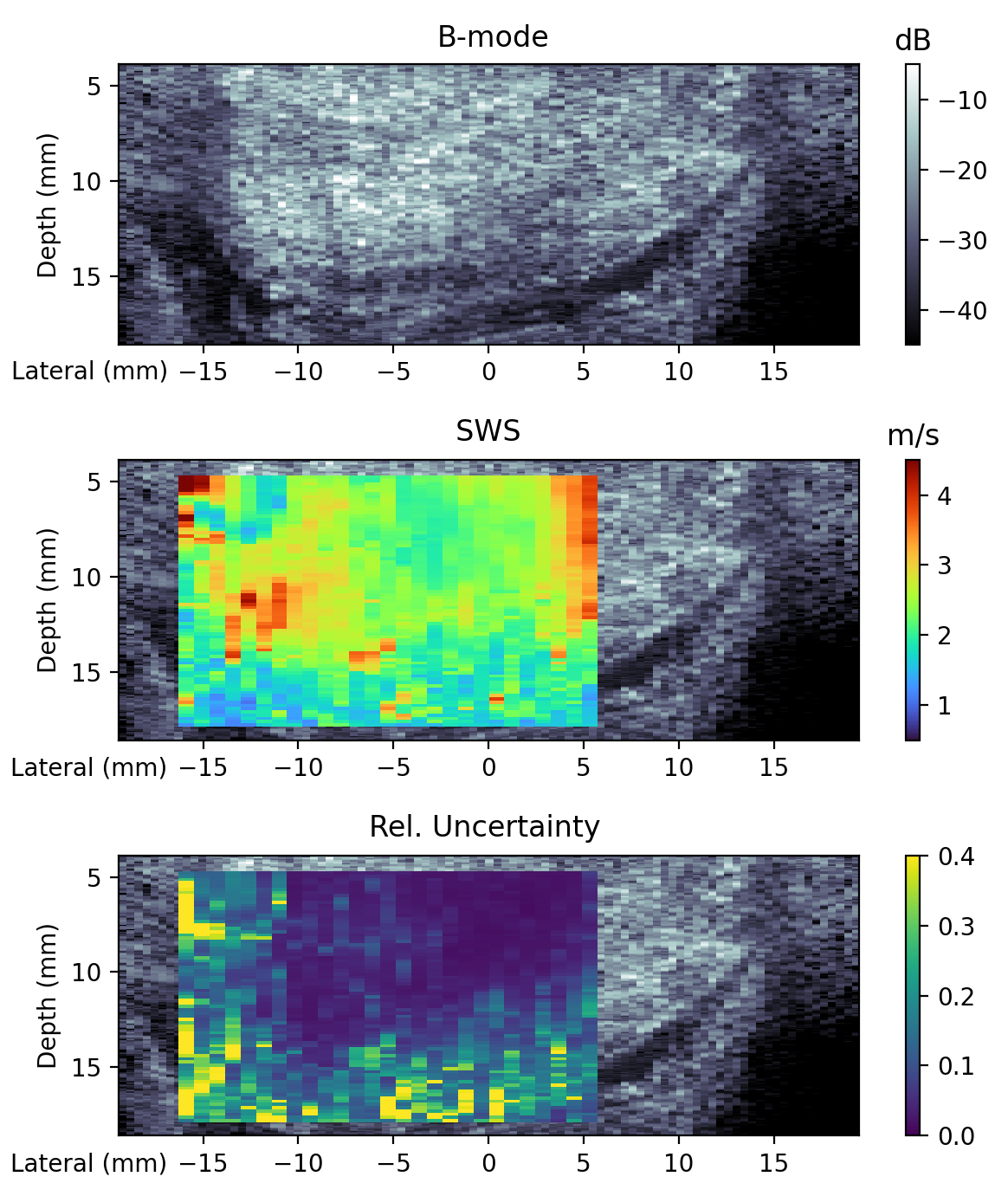}
	\caption{
	Example SWEI acquisition of the cervix showing the DNN-estimated SWSs and relative uncertainties for left-moving shear waves.
	Each vertical line in the SWS image corresponds to one of the 27 pushes in the acquisition.
	Relative uncertainty is the unitless ratio of uncertainty to SWS.
	The uncertainty was greater at the edges of the imaging window and near the cervical canal.
	}
	\label{fig:example}
\end{figure}
\begin{figure}[t]
	\centering
	\includegraphics[width=\columnwidth]{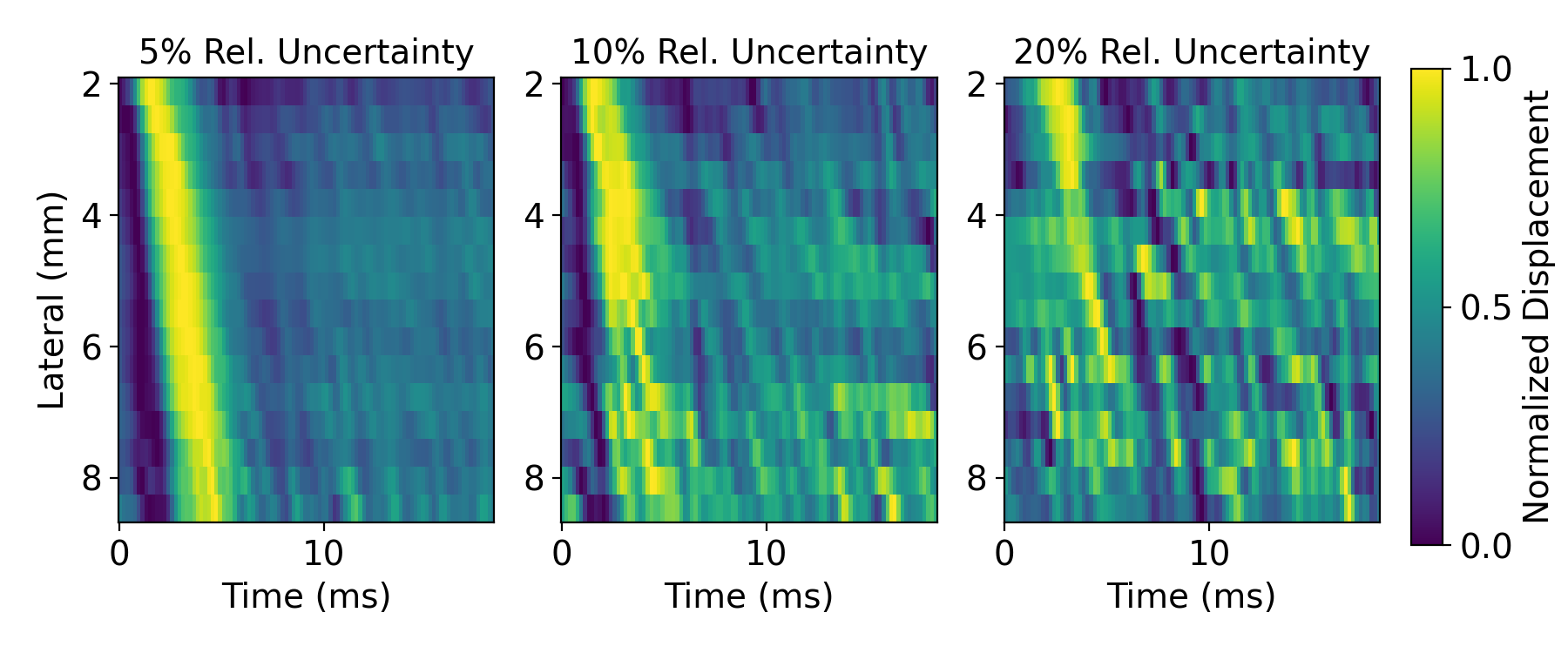}
	\caption{
	Three example space-time plots of data with 5\%, 10\%, and 20\%  relative uncertainty predicted by the model.
	The qualitative signal-to-noise ratio worsens with greater uncertainty.
	}
	\label{fig:quality}
\end{figure}
\begin{figure}[t]
	\centering
	\includegraphics[width=\columnwidth]{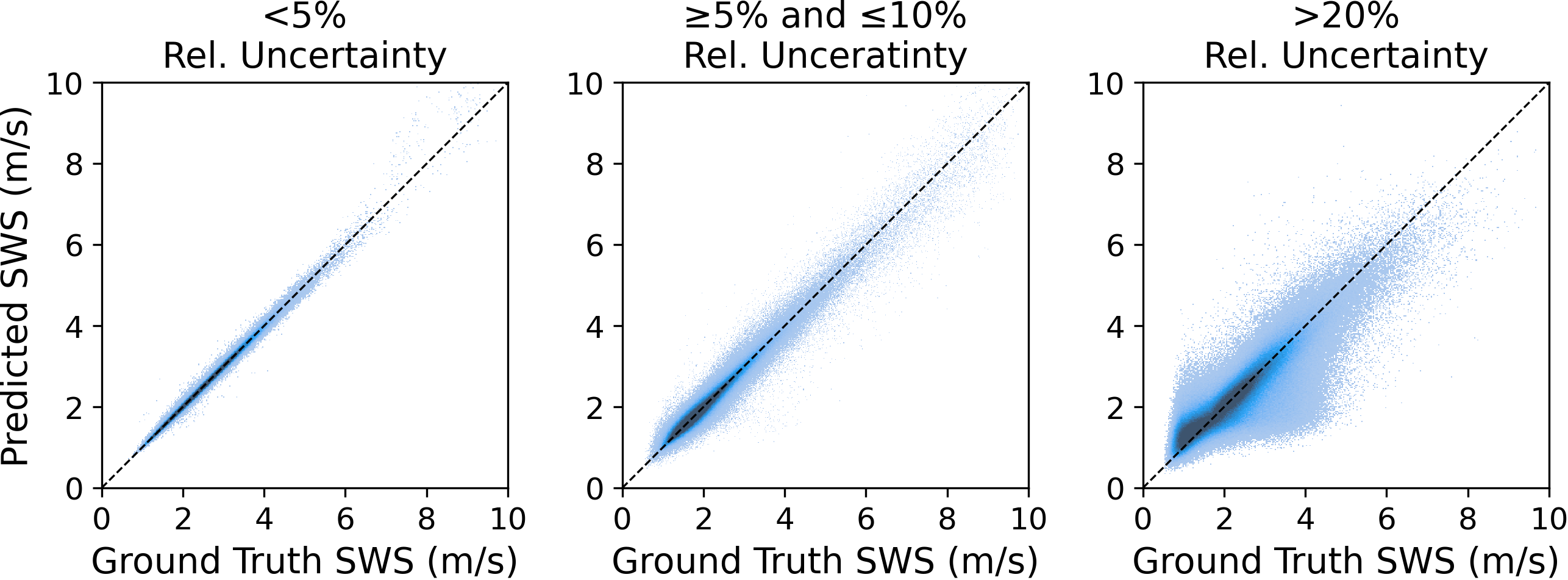}
	\caption{
	The distribution of ground truth versus predicted SWS for three ranges of relative uncertainty: less than 5\%, between 5\% and 10\%, and greater than 20\%.
	}
	\label{fig:yz}
\end{figure}

\subsection{External Dataset Testing}\label{sec:external}
We tested the trained model on an external dataset acquired by a collaborator in an \textit{ex vivo} stiffness phantom.
The isotropic, homogeneous phantom had a ground truth SWS of 5.13 m/s.
A Philips L7--4 linear transducer was used with a Verasonics Vantage 256 scanner.
A Siemens DAX curvilinear probe with a Siemens ACUSON Sequoia scanner was used to generate pushes through various pork belly body walls.
The dataset contained 929 independent space-time plots.

Here, we describe the additional data pre-processing steps necessary to apply the trained DNN to external data.
The primary challenges were: 1) that the spatiotemporal sampling rates differ from the cervix training data, 2) that the apparent SWSs are in different range than the training data, and 3) that particle velocity was used instead of particle displacement.

A different sampling rate for either the space or time dimensions changes the apparent SWS in a space-time plot if the difference is not accounted for.
The apparent SWS is
\begin{equation}
SWS_{apparent} = SWS_{true} \times \frac{\Delta x_0}{\Delta x} \times \frac{\Delta t}{\Delta t_0} \,,
\end{equation}
where $\Delta x$ and $\Delta t$ are spatial and temporal sampling intervals, and the subscript $0$ indicates the cervix training data.
The model estimates the apparent SWS, which must be converted to the true SWS by multiplying by the ratio $\dfrac{\Delta x}{\Delta t}\dfrac{\Delta t_0}{\Delta x_0}$.

The range of SWSs in the training set spanned 0.5--10.0~m/s, with an average of 2.1~m/s.
Model performance could be degraded if the apparent SWS is outside this range.
When an expected range is known, the input data can be interpolated to move the apparent SWS closer to 2.1~m/s and the model's output adjusted by the interpolation factor.
For example, in this external dataset, the expected SWS was around 5.0~m/s, so data were interpolated temporally by a factor of two in order to half the apparent SWS.

In the cervix study, particle displacement was used because the weak acoustic impulse of the small endovaginal transducer led to poor signal-to-noise in the velocity data.
However, particle velocity data is used in many SWE applications including in this external dataset.
The peak in particle velocity occurs earlier in time and is often followed by a significant negative trough.
Conversion of velocity to displacement data by integration leads to poor results due to accumulation of different DC offsets at each track location.
Instead, we can phase shift the velocity data by 90\textdegree\ to make the space-time plot resemble displacement data without the issues associated with accumulation.
This was done by taking the imaginary component of the Hilbert transform of the signal at each track location.
Fig.~\ref{fig:bofeng}A shows example SWE velocity data before and after phase shifting.

\section{Results}\label{sec:results}
\begin{figure}[t]
	\centering
	\begin{subfigure}[t]{0.48\columnwidth}
	\centering
	\includegraphics[width=\columnwidth]{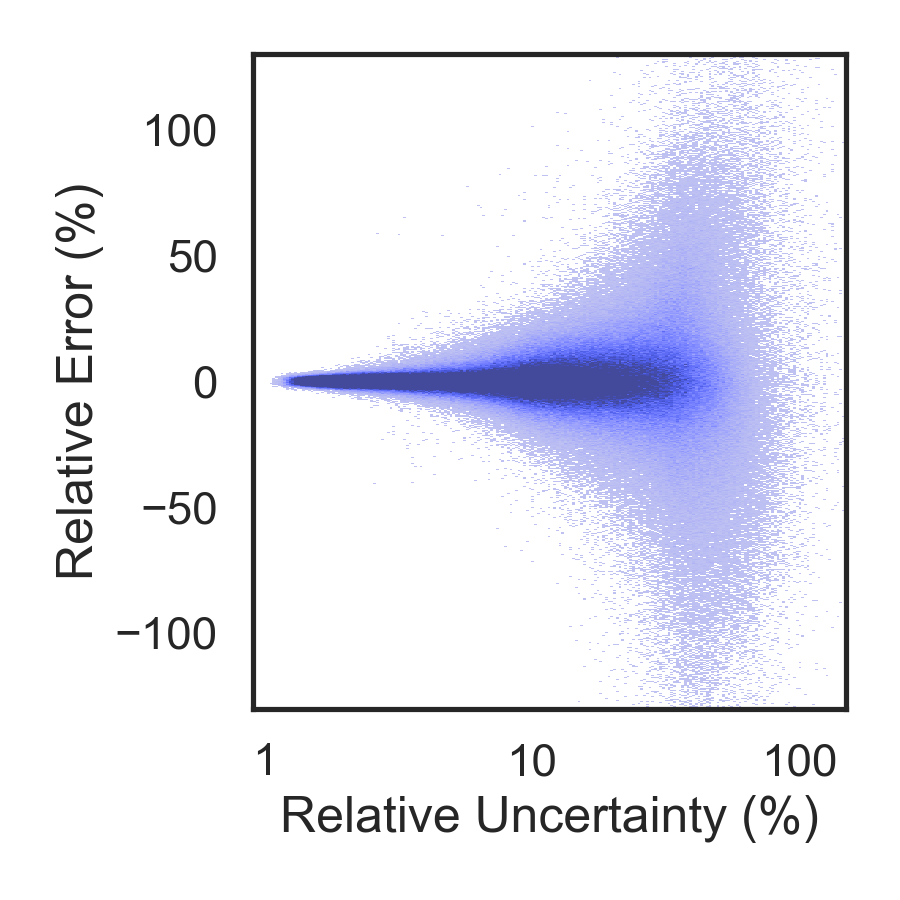}
	\caption{}
	\end{subfigure}
	\hfill
	\begin{subfigure}[t]{0.48\columnwidth}
	\centering
	\includegraphics[width=\columnwidth]{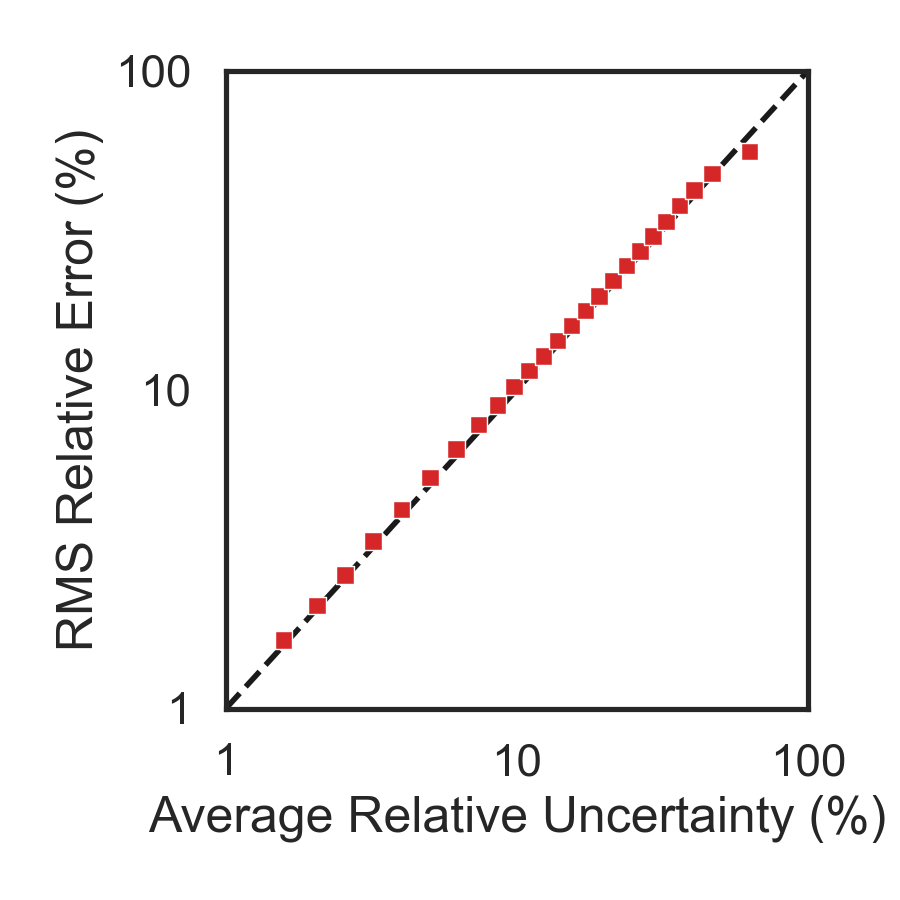}
	\caption{}
	\end{subfigure}
	\caption{
	(A) Plot of the relative uncertainty versus relative error.
	Inference was performed over the 29-patient dataset using leave-one-out cross-validation.
	(B) Points were grouped by uncertainty into 25 bins to compute the root-mean-square (RMS) relative error.
	Predicted uncertainty closely matched the observed RMS error.
	}
	\label{fig:calibration}
\end{figure}
Fig.~\ref{fig:example} shows the estimated SWS and uncertainty for an example 2D SWE acquisition in the cervix.
Only the left-moving shear waves are shown.
2000 space-time plots were processed to generate this image, with inference taking 2 seconds on a desktop CPU (Intel i7-6600U) or 50 ms on a desktop GPU (NVIDIA RTX 2080).

Fig.~\ref{fig:quality} shows data from three locations in Fig.~\ref{fig:example} with different levels of predicted uncertainty.
Higher uncertainty correlated with qualitatively worse signal-to-noise ratio and shear wave appearance in the space-time plot.

LOO cross-validation was used to perform inference on the entire 29-patient dataset.
Fig.~\ref{fig:yz} plots the predicted SWS versus ground truth SWS, stratified into three groups by level of uncertainty.
Fig.~\ref{fig:calibration}A plots the uncertainty versus error for all samples.
Points with greater estimated uncertainty were associated with a greater spread in true error.
Fig.~\ref{fig:calibration}B is the corresponding calibration plot, which shows estimated uncertainty strongly correlated with binned RMS error and each bin lying close to the identity line.
The mean absolute percent deviation for uncertainty calibration was 3.84\%.

\begin{figure}[!t]
	\centering
	\begin{subfigure}[b]{0.48\columnwidth}
	\centering
	\includegraphics[width=\columnwidth]{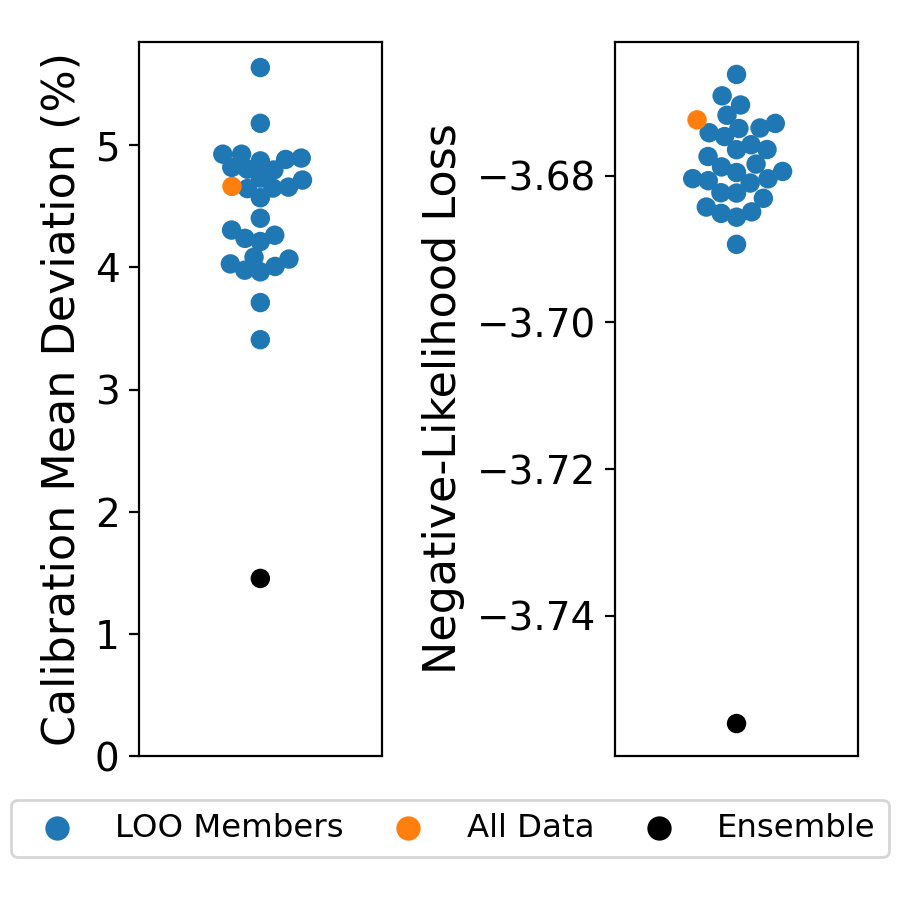}
	\caption{}
	\end{subfigure}
	\hfill
	\begin{subfigure}[b]{0.48\columnwidth}
	\centering
	\includegraphics[width=\columnwidth]{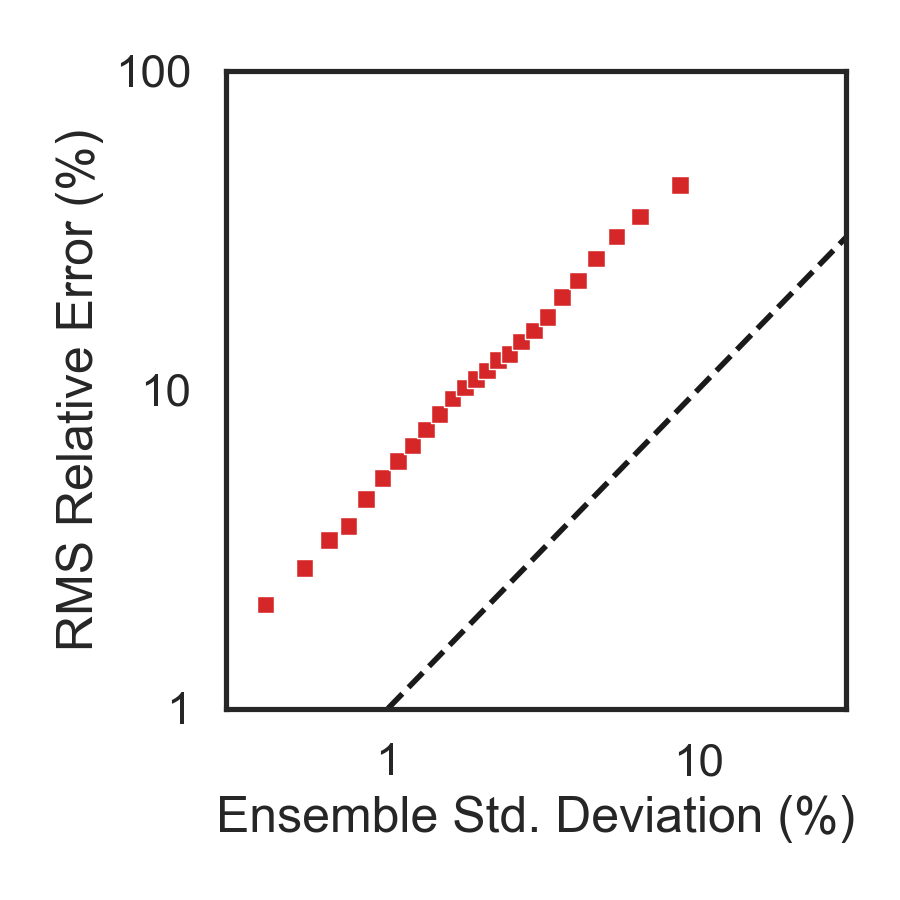}
	\caption{}
	\end{subfigure}
	\caption{
	(A) Plots of the calibration and average loss on held-out testing data. 
	We evaluated three cases: the 29 networks trained with leave-one-out (LOO) cross-validation [blue dots], a network trained on data from all 29 patients [orange dot], and the ensemble mean of the 29 LOO models [black dot].
	The ensemble had superior performance but takes 29 times longer to run.
	(B) A calibration curve where the standard deviation of the ensemble members' SWS estimates is used as a proxy for uncertainty.
	This way of quantifying uncertainty consistently underestimated the true RMS error.
	}
	\label{fig:ensemble}
\end{figure}
\subsection{Ensembles}
LOO training produced 29 sets of DNN weights, each trained with one patient's data left out.
We also trained a DNN using data from all 29 patients.
Additionally, we evaluated an ensemble model that uses the mean $\mu$ and mean $\sigma$ of all 29 LOO models.
To determine the best network configuration to use for inference in future applications, we evaluated these networks on data from the remaining patient that was set aside for testing.
Fig.~\ref{fig:ensemble}A plots the calibration mean deviation and the average  loss (equation~\ref{eq:lognorm_loss}) for these different models.
On the held-out test data, the ensemble model had better uncertainty calibration, with a mean deviation of 1.45\%, compared to any single network, with calibrations ranging from 3-6\%.

We also wondered whether the variance of the ensemble members' SWS estimates could be used as a proxy for uncertainty, thereby avoiding the need to train a model to explicitly predict uncertainty.
Fig.~\ref{fig:ensemble}B shows a calibration curve on the held-out testing data using the standard deviation of the SWS estimates as the uncertainty.
Although this quantity correlated with the true RMS error, it was not calibrated and underestimated the true RMS error by an average of 81\%, a factor of five.

\subsection{External Dataset}
The data pre-processing steps described in Section~\ref{sec:external} allowed us to apply the trained DNN model to the external dataset.
The median estimated SWS was 5.12 m/s, close to the ground truth value of 5.13 m/s.
Using velocity data without the phase-shifting step resulted in a positive bias (+0.24 m/s).
This bias was reduced for the phase-shifted data (--0.01 m/s).

\begin{figure}[t]
	\centering
	\begin{subfigure}[t]{0.48\columnwidth}
	\centering
	\includegraphics[width=\columnwidth]{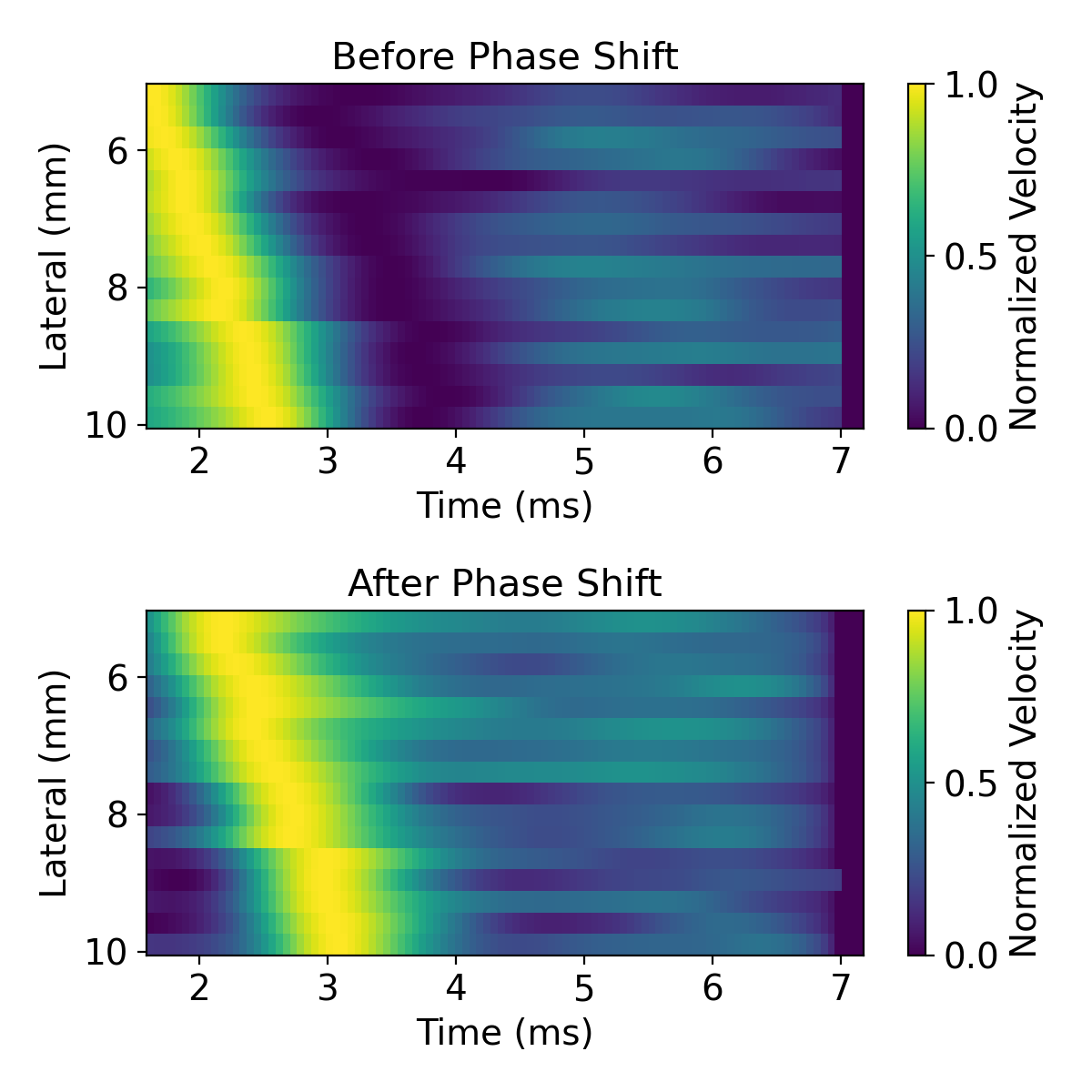}
	\caption{}
	\end{subfigure}
	\hfill
	\begin{subfigure}[t]{0.48\columnwidth}
	\centering
	\includegraphics[width=\columnwidth]{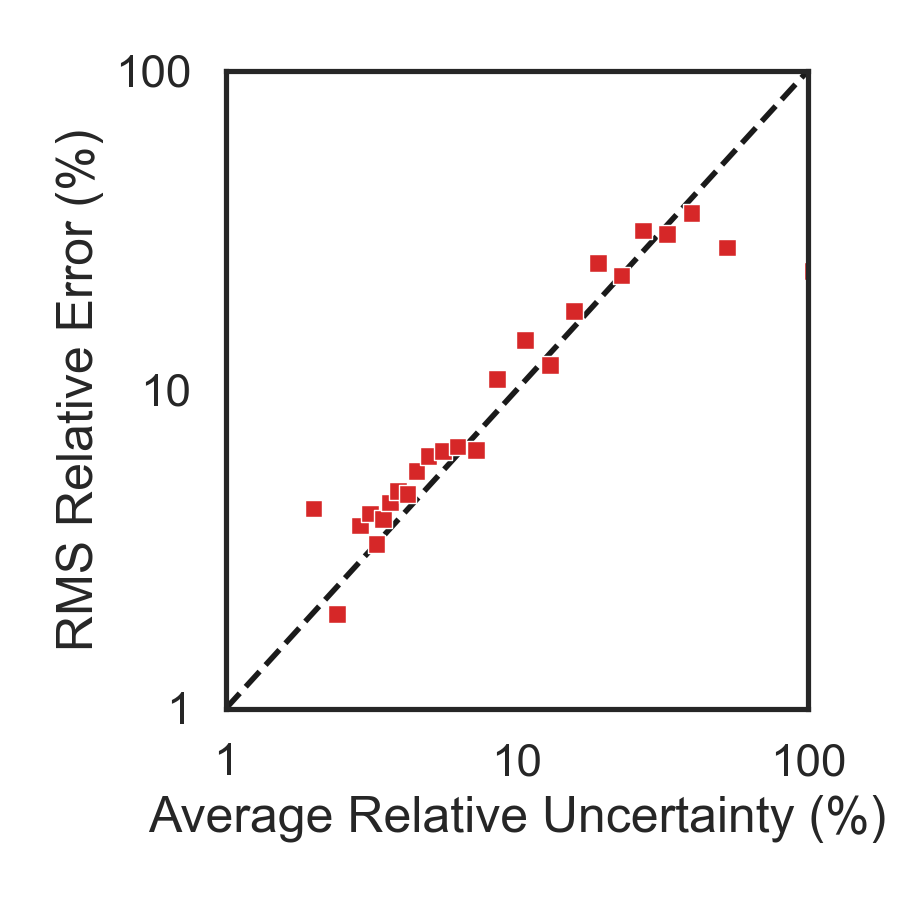}
	\caption{}
	\end{subfigure}
	\caption{
	(A) Sample of the external data collected with an L7--4 transducer and Verasonics scanner. 
	The normalized particle velocity data are shown before and after Hilbert transform-based phase shift, which transforms velocity data into the appearance of displacement data (like in Fig.~\ref{fig:spacetime}).
	(B) Calibration curve of the trained DNN applied on the external dataset.
	Points are close to but not directly on the identity line.
	RMS error increases with increasing uncertainty.
	}
	\label{fig:bofeng}
\end{figure}

The calibration curve of the estimated uncertainty is shown in Fig.~\ref{fig:bofeng}B.
The mean absolute percent deviation was 29.7\%.

\section{Discussion}\label{sec:discussion}
Using a large \invivo SWE dataset, we trained a deep learning model to simultaneously perform SWS estimation and provide a quantitative estimate of the predictive uncertainty.
We now call this model ``SweiNet''.
While existing SWS estimators are reliable and perform well, some are too slow, and none provides an easily interpretable or calibrated uncertainty metric.
SweiNet runs in real time and provides a well-calibrated uncertainty that closely approximates the true spread in predictive error.

\subsection{Processing Speed}
In point-SWE, the SWS is calculated for one right-moving and one left-moving shear wave to produce the point stiffness measurement.
2D SWE is much more computationally expensive because the SWS needs to be estimated for a large number of points in the ROI to produce a stiffness map.
A typical 2D SWE acquisition from the cervix dataset consisted of 27 push events, $\sim$80 axial locations, and 2 wave propagation directions: this totals over 4000 space-time plots per stiffness map.

The RANSAC algorithm is robust to outliers, and the percent inliers correlates with data quality.
Unfortunately, RANSAC is not suitable for real-time elastography because the algorithm is non-deterministic and computationally slow, a problem magnified when making 2D stiffness maps.
The algorithm can take up to 20 minutes to process data from a single acquisition.
Faster algorithms, like the Radon transform method, lack a useful uncertainty metric.

Our deep learning approach to SWS estimation is deterministic and non-iterative.
The inference time of SweiNet on a laptop CPU is 2.0 seconds and can be decreased further with GPU hardware acceleration.
Moving forward, we plan to implement SweiNet on a clinical ultrasound scanner to give sonographers real-time feedback about data quality.

\subsection{Uncertainty Quantification}
The uncertainty of any clinical measurement is important for its interpretation and use in decision-making.
Uncertainty metrics associated with existing SWS estimators are lacking and unsatisfactory.

Building upon recent deep learning research~\cite{nix1994estimating,kendall2017uncertainties,lakshminarayanan2017simple}, we trained our SweiNet model to output the two parameters $m$ and $\sigma$ a log-normal likelihood distribution for the SWS.
This approach learns the uncertainty calibration implicitly from the training process.
The most natural uncertainty metric is the unitless relative uncertainty $\sinh{\sigma}$, which may be multiplied by the estimated SWS to obtain an uncertainty in units of meters per second: $m \sinh{\sigma}$.

The uncertainty does not say whether any one SWS estimate is wrong.
Rather, it describes the degree of spread in error averaged over a sample of SWS estimates.
We quantified uncertainty calibration by grouping all SWS estimates into 25 bins of increasing uncertainty and comparing how well the average uncertainty matches the RMS error.
In Fig.~\ref{fig:calibration}B, every bin except for the final bin had an absolute deviation of less than 4\%.
SweiNet's uncertainty calibration was superior to that of the previous iteration we trained on point-SWE data~\cite{jin2021deep}.

The final bin, containing samples with the largest predicted relative uncertainty, overestimated the observed RMS relative error.
This may be because the ground truth SWS estimates were constrained between 0.5 and 12 m/s, placing a cap on error.

\subsection{Uses for Uncertainty}
Beyond using uncertainty simply as a metric of data quality, access to a well-calibrated uncertainty opens many doors for downstream processing.
In particular, SweiNet describes a probability distribution, which can be operated on directly.
The distribution remains log-normal under power and scaling transformations, and the conversion to shear modulus via $G=\rho \, c^2$ is
\begin{equation}
\begin{split}
SWS &\sim \textrm{LogNormal}(m, \,\sigma) \\
G &\sim \textrm{LogNormal}(\rho m^2, \,2\sigma) \,.
\end{split}
\end{equation}
If one works with the log-domain variables, then these distributions are normal.

Given multiple SWS measurements, we may want to compute an average.
The variance of the weighted average of random variables is 
\begin{equation}
\sum_{i,j} w_i w_j \sigma_i \sigma_j \rho_{ij} \,,
\end{equation}
where $w$ are the weights, $\sigma^2$ are the variances, and $\rho_{ij}$ is the correlation coefficient.
If all measurements are independent, and thus uncorrelated, the variance of the weighted average simplifies to $\sum_i w_i^2 \sigma_i^2$.
This quantity is minimized when the weights are inversely proportional to the variances:
\begin{equation}\label{eq:weight}
w_i = \frac{1}{\sigma^2_i} \left( \sum_k \frac{1}{\sigma^2_k} \right)^{-1} \,.
\end{equation}
In a 2D SWEI acquisition, a SWS is calculated at each spatial location in the ROI; however, these measurements are highly spatially correlated and are \textit{not} independent.
For averaging the SWSs from a \textit{single} 2D SWEI acquisition, we recommend using the weights in equation~\ref{eq:weight} but assuming $\rho_{ij}=1$, such that the variance of the average is maximized.
The degree of correlation between nearby points depends on the distance between them, the data quality, and tissue heterogeneity.
Further analysis is needed to provide estimates of the spatial correlation of SWSs in \invivo data.

SWS probability distributions can also be used for maximum likelihood estimation of derived parameters.
For example, consider SWS data of skeletal muscle measured at various rotation angles relative to the muscle fibers and the process of fitting an ellipse to these data~\cite{knight2021full}.
Rather than optimizing for the smallest mean error, we could find a curve with the maximum joint likelihood of all SWS data points, using SweiNet's output as the probability function.
Points with high uncertainty have a large spread in their probability distribution and would contribute less to the fit.
This process decreases the influence of points with high uncertainty and could prevent them from skewing the estimated ellipse.

\subsection{Ensembles}
The model trained on all data had no better performance than the LOO models.
While initially surprising, its training set was only increased by one additional patient's data.
The ensemble of 29 networks had better uncertainty calibration and much lower loss than any other individual network.
This agrees with a large body of work showing that ensemble models generally outperform individual members.
Specifically, creating an ensemble of uncertainty-predicting neural networks was previously proposed in~\cite{lakshminarayanan2017simple}.
However, we found that their Gaussian-mixture formula for calculating the mean variance performed poorly, and we instead used the arithmetic mean for both $m$ and $\sigma$.

In considering whether to use a single network or an ensemble for inference, we must weigh the benefit in quality with the computational cost of running 29 networks.
For example, running the ensemble sequentially on a desktop CPU takes roughly 90 seconds to process all space-time plots in a single SWE acquisition:
this is prohibitively too long for real-time imaging.
A GPU reduces inference time significantly: ensemble inference takes 3.0 seconds on an NVIDIA Tesla T4 and 2.0 seconds on an NVIDIA RTX 2080.
The former GPU has performance capabilities similar to what might be equipped on an ultrasound scanner.

We also tested whether the standard deviation of SWS estimates from ensemble members was a good proxy for predictive uncertainty.
While this metric did correlate with true RMS error, it severely and consistently underestimated it.
This result shows that the spread in ensemble members' SWS estimates reflects uncertainty but is not calibrated.
Calibration is learned during training.

\subsection{Application to External Datasets}
In order for SweiNet to be useful beyond cervix SWE and our specific imaging configuration, it must generalize to external data.
Model performance, especially the uncertainty's calibration, would likely be poor if the input appears substantially different from the training data.
We addressed three challenges in adapting the model to external data: different spatial and temporal sampling rates, different range of apparent SWSs, and using particle velocity instead of displacement.

In particular, velocity data typically has a strong peak and trough whereas displacement data usually has no or a weak trough.
We used the Hilbert transform to phase shift velocity data, making it appear like displacement data while avoiding accumulation-related artifacts.
We have not seen this approach reported in the literature, possibly because traditional algorithms are not sensitive to the difference in data appearance.

We found that the predicted uncertainty metric was a good estimate for the true RMS error, but uncertainty calibration was worse than with the cervix data.
However, points remained clustered near the identity line, indicating neither a severe overestimation nor underestimation of uncertainty.

\subsection{Limitations}
In determining how to quantify uncertainty for ultrasound SWE, we must first consider what sources of uncertainty exist and which are observable.
The deep learning model in this paper takes the tracked displacement data as input to estimate SWS and uncertainty.
Therefore, SweiNet is only sensitive to factors that are observable in the displacement data and that affect the likelihood of SWS estimation error.
These factors include signal-to-noise ratio, tracking artifacts, tissue heterogeneity, attenuation/dispersion, and speckle bias.
These sources of error can all leave a signature in the space-time plot.

A number of factors may effect error without leaving any sign in the space-time plot.
SweiNet estimates the apparent SWS, which under some conditions would not be equal to the true SWS.
For example:
1) tissue compression, often by an operator pressing too hard with the transducer, increases the apparent SWS;
2) non-alignment of the shear wavefront with the tracking direction increases the apparent SWS~\cite{song2014fast}.

SweiNet's learned uncertainty is calibrated to the cervix data it was trained on.
The uncertainty was still somewhat calibrated when SweiNet was applied to the external data acquired with a different transducer and scanner. 
However, we believe that, in general, the uncertainty would not be well-calibrated on external data because the degree to which factors like noise and artifacts affect the likelihood of SWS estimation error should depend on the imaging setup and the organ being imaged.
Fine-tuning could help re-calibrate the trained model to different domains.

\section{Conclusion}
Ultrasound SWE imaging is a low-cost, non-invasive technique that  quantitatively assesses tissue stiffness by measuring the SWS.
We have developed a deep learning model called SweiNet to simultaneously perform SWS estimation and output a quantitative well-calibrated uncertainty metric.
We are prospectively testing SweiNet in an ongoing clinical trial investigating cervical stiffness and induction of labor outcomes.
We are also working to implement the model on an ultrasound scanner to provide real-time SWS and uncertainty estimates at the bedside.
Access to calibrated uncertainty values could help clinicians better interpret SWS measurements and facilitate downstream applications of SWE.
We have made the code and trained model publicly available as a ready-to-use tool and to aid members of the research community in training their own models.

\appendices
\section*{Acknowledgment}
We would like to thank Dr.\ Bofeng Zhang for providing the external SWE data.
We would also like to thank Dr.\ Ivan Rosado-Mendez for his helpful discussions.
This work was supported by NIH grants T32GM007171, R01HD072077, R01HD096361.
Our laboratory receives in-kind support from Siemens Healthineers, Ultrasound Division.
M.~Palmeri holds intellectual property related to acoustic radiation force impulse and shear wave elasticity imaging.
H.~Feltovich and T.~Hall hold intellectual property related to quantitative ultrasound methods for cervical assessment.

\bibliographystyle{IEEEtran}
\bibliography{IEEEabrv,references}

\begin{thebibliography}{10}
\providecommand{\url}[1]{#1}
\csname url@samestyle\endcsname
\providecommand{\newblock}{\relax}
\providecommand{\bibinfo}[2]{#2}
\providecommand{\BIBentrySTDinterwordspacing}{\spaceskip=0pt\relax}
\providecommand{\BIBentryALTinterwordstretchfactor}{4}
\providecommand{\BIBentryALTinterwordspacing}{\spaceskip=\fontdimen2\font plus
\BIBentryALTinterwordstretchfactor\fontdimen3\font minus
  \fontdimen4\font\relax}
\providecommand{\BIBforeignlanguage}[2]{{%
\expandafter\ifx\csname l@#1\endcsname\relax
\typeout{** WARNING: IEEEtran.bst: No hyphenation pattern has been}%
\typeout{** loaded for the language `#1'. Using the pattern for}%
\typeout{** the default language instead.}%
\else
\language=\csname l@#1\endcsname
\fi
#2}}
\providecommand{\BIBdecl}{\relax}
\BIBdecl

\bibitem{sarvazyan1998shear}
A.~P. Sarvazyan, O.~V. Rudenko, S.~D. Swanson, J.~B. Fowlkes, and S.~Y.
  Emelianov, ``Shear wave elasticity imaging: a new ultrasonic technology of
  medical diagnostics,'' \emph{Ultrasound in medicine \& biology}, vol.~24,
  no.~9, pp. 1419--1435, 1998.

\bibitem{sarvazyan2011overview}
A.~Sarvazyan, T.~J~Hall, M.~W~Urban, M.~Fatemi, S.~R~Aglyamov, and B.~S~Garra,
  ``An overview of elastography-an emerging branch of medical imaging,''
  \emph{Current Medical Imaging}, vol.~7, no.~4, pp. 255--282, 2011.

\bibitem{nightingale2003shear}
K.~Nightingale, S.~McAleavey, and G.~Trahey, ``Shear-wave generation using
  acoustic radiation force: in vivo and ex vivo results,'' \emph{Ultrasound in
  medicine \& biology}, vol.~29, no.~12, pp. 1715--1723, 2003.

\bibitem{kasai1985real}
C.~Kasai, K.~Namekawa, A.~Koyano, and R.~Omoto, ``Real-time two-dimensional
  blood flow imaging using an autocorrelation technique,'' \emph{IEEE
  Transactions on sonics and ultrasonics}, vol.~32, no.~3, pp. 458--464, 1985.

\bibitem{loupas1995axial}
T.~Loupas, J.~Powers, and R.~W. Gill, ``An axial velocity estimator for
  ultrasound blood flow imaging, based on a full evaluation of the doppler
  equation by means of a two-dimensional autocorrelation approach,'' \emph{IEEE
  transactions on ultrasonics, ferroelectrics, and frequency control}, vol.~42,
  no.~4, pp. 672--688, 1995.

\bibitem{pinton2006rapid}
G.~F. Pinton, J.~J. Dahl, and G.~E. Trahey, ``Rapid tracking of small
  displacements with ultrasound,'' \emph{IEEE transactions on ultrasonics,
  ferroelectrics, and frequency control}, vol.~53, no.~6, pp. 1103--1117, 2006.

\bibitem{wang2010improving}
M.~H. Wang, M.~L. Palmeri, V.~M. Rotemberg, N.~C. Rouze, and K.~R. Nightingale,
  ``Improving the robustness of time-of-flight based shear wave speed
  reconstruction methods using ransac in human liver in vivo,''
  \emph{Ultrasound in medicine \& biology}, vol.~36, no.~5, pp. 802--813, 2010.

\bibitem{rouze2012parameters}
N.~C. Rouze, M.~H. Wang, M.~L. Palmeri, and K.~R. Nightingale, ``Parameters
  affecting the resolution and accuracy of 2-d quantitative shear wave
  images,'' \emph{IEEE transactions on ultrasonics, ferroelectrics, and
  frequency control}, vol.~59, no.~8, pp. 1729--1740, 2012.

\bibitem{mcaleavey2015shear}
S.~A. McAleavey, L.~O. Osapoetra, and J.~Langdon, ``Shear wave arrival time
  estimates correlate with local speckle pattern,'' \emph{IEEE transactions on
  ultrasonics, ferroelectrics, and frequency control}, vol.~62, no.~12, pp.
  2054--2067, 2015.

\bibitem{deng2016system}
Y.~Deng, N.~C. Rouze, M.~L. Palmeri, and K.~R. Nightingale, ``On
  system-dependent sources of uncertainty and bias in ultrasonic quantitative
  shear-wave imaging,'' \emph{IEEE Trans Ultrason Ferroelectr Freq Control},
  vol.~63, no.~3, pp. 381--93, 2016.

\bibitem{palmeri2008quantifying}
M.~L. Palmeri, M.~H. Wang, J.~J. Dahl, K.~D. Frinkley, and K.~R. Nightingale,
  ``Quantifying hepatic shear modulus in vivo using acoustic radiation force,''
  \emph{Ultrasound in medicine \& biology}, vol.~34, no.~4, pp. 546--558, 2008.

\bibitem{tanter2008quantitative}
M.~Tanter, J.~Bercoff, A.~Athanasiou, T.~Deffieux, J.-L. Gennisson,
  G.~Montaldo, M.~Muller, A.~Tardivon, and M.~Fink, ``Quantitative assessment
  of breast lesion viscoelasticity: initial clinical results using supersonic
  shear imaging,'' \emph{Ultrasound in medicine \& biology}, vol.~34, no.~9,
  pp. 1373--1386, 2008.

\bibitem{song2014fast}
P.~Song, A.~Manduca, H.~Zhao, M.~W. Urban, J.~F. Greenleaf, and S.~Chen, ``Fast
  shear compounding using robust 2-d shear wave speed calculation and
  multi-directional filtering,'' \emph{Ultrasound in medicine \& biology},
  vol.~40, no.~6, pp. 1343--1355, 2014.

\bibitem{rouze2010robust}
N.~C. Rouze, M.~H. Wang, M.~L. Palmeri, and K.~R. Nightingale, ``Robust
  estimation of time-of-flight shear wave speed using a radon sum
  transformation,'' \emph{IEEE transactions on ultrasonics, ferroelectrics, and
  frequency control}, vol.~57, no.~12, pp. 2662--2670, 2010.

\bibitem{urban2012use}
M.~W. Urban and J.~F. Greenleaf, ``Use of the radon transform for estimation of
  shear wave speed,'' \emph{The Journal of the Acoustical Society of America},
  vol. 132, no.~3, pp. 1982--1982, 2012.

\bibitem{jin2022radon}
\BIBentryALTinterwordspacing
F.~Jin, ``A radon transform wave-speed estimator,'' Mar. 2022. [Online].
  Available: \url{https://doi.org/10.5281/zenodo.6364164}
\BIBentrySTDinterwordspacing

\bibitem{kijanka2018local}
P.~Kijanka and M.~W. Urban, ``Local phase velocity based imaging: A new
  technique used for ultrasound shear wave elastography,'' \emph{IEEE
  transactions on medical imaging}, vol.~38, no.~4, pp. 894--908, 2018.

\bibitem{deng2016ultrasonic}
Y.~Deng, N.~C. Rouze, M.~L. Palmeri, and K.~R. Nightingale, ``Ultrasonic shear
  wave elasticity imaging sequencing and data processing using a verasonics
  research scanner,'' \emph{IEEE transactions on ultrasonics, ferroelectrics,
  and frequency control}, vol.~64, no.~1, pp. 164--176, 2016.

\bibitem{rouze2018characterization}
N.~C. Rouze, Y.~Deng, C.~A. Trutna, M.~L. Palmeri, and K.~R. Nightingale,
  ``Characterization of viscoelastic materials using group shear wave speeds,''
  \emph{IEEE transactions on ultrasonics, ferroelectrics, and frequency
  control}, vol.~65, no.~5, pp. 780--794, 2018.

\bibitem{lecun2015deep}
Y.~LeCun, Y.~Bengio, and G.~Hinton, ``Deep learning,'' \emph{nature}, vol. 521,
  no. 7553, pp. 436--444, 2015.

\bibitem{luchies2018deep}
A.~C. Luchies and B.~C. Byram, ``Deep neural networks for ultrasound
  beamforming,'' \emph{IEEE transactions on medical imaging}, vol.~37, no.~9,
  pp. 2010--2021, 2018.

\bibitem{hyun2019beamforming}
D.~Hyun, L.~L. Brickson, K.~T. Looby, and J.~J. Dahl, ``Beamforming and speckle
  reduction using neural networks,'' \emph{IEEE transactions on ultrasonics,
  ferroelectrics, and frequency control}, vol.~66, no.~5, pp. 898--910, 2019.

\bibitem{chan2021deep}
D.~Y. Chan, D.~C. Morris, T.~J. Polascik, M.~L. Palmeri, and K.~R. Nightingale,
  ``Deep convolutional neural networks for displacement estimation in arfi
  imaging,'' \emph{IEEE Transactions on Ultrasonics, Ferroelectrics, and
  Frequency Control}, vol.~68, no.~7, pp. 2472--2481, 2021.

\bibitem{tehrani2020displacement}
A.~K. Tehrani and H.~Rivaz, ``Displacement estimation in ultrasound
  elastography using pyramidal convolutional neural network,'' \emph{IEEE
  Transactions on Ultrasonics, Ferroelectrics, and Frequency Control}, 2020.

\bibitem{ahmed2020dswe}
S.~Ahmed, U.~Kamal, and M.~K. Hasan, ``Dswe-net: A deep learning approach for
  shear wave elastography and lesion segmentation using single push acoustic
  radiation force,'' \emph{Ultrasonics}, p. 106283, 2020.

\bibitem{hein2019relu}
M.~Hein, M.~Andriushchenko, and J.~Bitterwolf, ``Why relu networks yield
  high-confidence predictions far away from the training data and how to
  mitigate the problem,'' in \emph{Proceedings of the IEEE Conference on
  Computer Vision and Pattern Recognition}, 2019, pp. 41--50.

\bibitem{nguyen2015deep}
A.~Nguyen, J.~Yosinski, and J.~Clune, ``Deep neural networks are easily fooled:
  High confidence predictions for unrecognizable images,'' in \emph{Proceedings
  of the IEEE conference on computer vision and pattern recognition}, 2015, pp.
  427--436.

\bibitem{ovadia2019can}
Y.~Ovadia, E.~Fertig, J.~Ren, Z.~Nado, D.~Sculley, S.~Nowozin, J.~Dillon,
  B.~Lakshminarayanan, and J.~Snoek, ``Can you trust your model's uncertainty?
  evaluating predictive uncertainty under dataset shift,'' in \emph{Advances in
  Neural Information Processing Systems}, 2019, pp. 13\,991--14\,002.

\bibitem{nix1994estimating}
D.~A. Nix and A.~S. Weigend, ``Estimating the mean and variance of the target
  probability distribution,'' in \emph{Proceedings of 1994 ieee international
  conference on neural networks (ICNN'94)}, vol.~1.\hskip 1em plus 0.5em minus
  0.4em\relax IEEE, 1994, pp. 55--60.

\bibitem{kendall2017uncertainties}
A.~Kendall and Y.~Gal, ``What uncertainties do we need in bayesian deep
  learning for computer vision?'' in \emph{Advances in neural information
  processing systems}, 2017, pp. 5574--5584.

\bibitem{gal2016dropout}
Y.~Gal and Z.~Ghahramani, ``Dropout as a bayesian approximation: Representing
  model uncertainty in deep learning,'' in \emph{international conference on
  machine learning}, 2016, pp. 1050--1059.

\bibitem{lakshminarayanan2017simple}
B.~Lakshminarayanan, A.~Pritzel, and C.~Blundell, ``Simple and scalable
  predictive uncertainty estimation using deep ensembles,'' in \emph{Advances
  in neural information processing systems}, 2017, pp. 6402--6413.

\bibitem{li2016learning}
C.~Li, A.~Stevens, C.~Chen, Y.~Pu, Z.~Gan, and L.~Carin, ``Learning weight
  uncertainty with stochastic gradient mcmc for shape classification,'' in
  \emph{Proceedings of the IEEE Conference on Computer Vision and Pattern
  Recognition}, 2016, pp. 5666--5675.

\bibitem{zhang2019cyclical}
R.~Zhang, C.~Li, J.~Zhang, C.~Chen, and A.~G. Wilson, ``Cyclical stochastic
  gradient mcmc for bayesian deep learning,'' \emph{arXiv preprint
  arXiv:1902.03932}, 2019.

\bibitem{timmons2010cervical}
B.~Timmons, M.~Akins, and M.~Mahendroo, ``Cervical remodeling during pregnancy
  and parturition,'' \emph{Trends in Endocrinology \& Metabolism}, vol.~21,
  no.~6, pp. 353--361, 2010.

\bibitem{feltovich2019labour}
H.~Feltovich, ``Labour and delivery: a clinician's perspective on a
  biomechanics problem,'' \emph{Interface focus}, vol.~9, no.~5, p. 20190032,
  2019.

\bibitem{carlson2019quantitative}
L.~C. Carlson, T.~J. Hall, I.~M. Rosado-Mendez, L.~Mao, and H.~Feltovich,
  ``Quantitative assessment of cervical softening during pregnancy with shear
  wave elasticity imaging: an in vivo longitudinal study,'' \emph{Interface
  focus}, vol.~9, no.~5, p. 20190030, 2019.

\bibitem{swiatkowska2011elastography}
M.~Swiatkowska-Freund and K.~Preis, ``Elastography of the uterine cervix:
  implications for success of induction of labor,'' \emph{Ultrasound in
  Obstetrics \& Gynecology}, vol.~38, no.~1, pp. 52--56, 2011.

\bibitem{wang2019diagnostic}
B.~Wang, Y.~Zhang, S.~Chen, X.~Xiang, J.~Wen, M.~Yi, B.~He, and B.~Hu,
  ``Diagnostic accuracy of cervical elastography in predicting preterm
  delivery: A systematic review and meta-analysis,'' \emph{Medicine}, vol.~98,
  no.~29, 2019.

\bibitem{lu2020predictive}
J.~Lu, Y.~K.~Y. Cheng, S.~Y.~S. Ho, D.~S. Sahota, L.~Hui, L.~C. Poon, and T.~Y.
  Leung, ``The predictive value of cervical shear wave elastography in the
  outcome of labor induction,'' \emph{Acta obstetricia et gynecologica
  Scandinavica}, vol.~99, no.~1, pp. 59--68, 2020.

\bibitem{jin2021deep}
F.~Q. Jin, L.~C. Carlson, T.~J. Hall, H.~Feltovich, and M.~L. Palmeri, ``Deep
  learning based quantitative uncertainty estimation for ultrasound shear wave
  elasticity imaging,'' in \emph{2021 IEEE International Ultrasonics Symposium
  (IUS)}.\hskip 1em plus 0.5em minus 0.4em\relax IEEE, 2021, pp. 1--4.

\bibitem{doherty2013harmonic}
J.~R. Doherty, J.~J. Dahl, and G.~E. Trahey, ``Harmonic tracking of acoustic
  radiation force-induced displacements,'' \emph{IEEE transactions on
  ultrasonics, ferroelectrics, and frequency control}, vol.~60, no.~11, pp.
  2347--2358, 2013.

\bibitem{lipman2016evaluating}
S.~L. Lipman, N.~C. Rouze, M.~L. Palmeri, and K.~R. Nightingale, ``Evaluating
  the improvement in shear wave speed image quality using multidimensional
  directional filters in the presence of reflection artifacts,'' \emph{IEEE
  transactions on ultrasonics, ferroelectrics, and frequency control}, vol.~63,
  no.~8, pp. 1049--1063, 2016.

\bibitem{he2016identity}
K.~He, X.~Zhang, S.~Ren, and J.~Sun, ``Identity mappings in deep residual
  networks,'' in \emph{European conference on computer vision}.\hskip 1em plus
  0.5em minus 0.4em\relax Springer, 2016, pp. 630--645.

\bibitem{smith2019super}
L.~N. Smith and N.~Topin, ``Super-convergence: Very fast training of neural
  networks using large learning rates,'' in \emph{Artificial Intelligence and
  Machine Learning for Multi-Domain Operations Applications}, vol. 11006.\hskip
  1em plus 0.5em minus 0.4em\relax International Society for Optics and
  Photonics, 2019, p. 1100612.

\bibitem{wang2013precision}
M.~Wang, B.~Byram, M.~Palmeri, N.~Rouze, and K.~Nightingale, ``On the precision
  of time-of-flight shear wave speed estimation in homogeneous soft solids:
  initial results using a matrix array transducer,'' \emph{IEEE transactions on
  ultrasonics, ferroelectrics, and frequency control}, vol.~60, no.~4, pp.
  758--770, 2013.

\bibitem{knight2021full}
A.~E. Knight, C.~A. Trutna, N.~C. Rouze, L.~D. Hobson-Webb, A.~Caenen, F.~Q.
  Jin, M.~L. Palmeri, and K.~R. Nightingale, ``Full characterization of in vivo
  muscle as an elastic, incompressible, transversely isotropic material using
  ultrasonic rotational 3d shear wave elasticity imaging,'' \emph{IEEE
  Transactions on Medical Imaging}, vol.~41, no.~1, pp. 133--144, 2021.

\end{thebibliography}

\end{document}